\documentclass[pdflatex,sn-nature]{sn-jnl}

\usepackage{graphicx}%
\usepackage{multirow}%
\usepackage{amsmath,amssymb,amsfonts}%
\usepackage{amsthm}%
\usepackage{bm}
\usepackage{mathrsfs}%
\usepackage[title]{appendix}%
\usepackage{xcolor}%
\usepackage{textcomp}%
\usepackage{manyfoot}%
\usepackage{booktabs}%
\usepackage{listings}%
\usepackage{indentfirst}

\begin{document}

\title[Article Title]{
Direct demonstration of electric chirality control in a helimagnetic YMn$_6$Sn$_6$ by spin-polarized neutron scattering
}

\author*[1] {\fnm{Hidetoshi} \sur{Masuda}} \email{hidetoshi.masuda.c8@tohoku.ac.jp}

\author[1]  {\fnm{Yutaro} \sur{Yanagisawa}}

\author[2]  {\fnm{Kazuki} \sur{Ohishi}}

\author[3]  {\fnm{Yusuke} \sur{Nambu}}

\author[4]  {\fnm{Yoichi} \sur{Nii}}

\author[1,5]{\fnm{Yoshinori} \sur{Onose}}

\affil[1]{\orgdiv{Institute for Materials Research}, 
\orgname{Tohoku University}, 
\orgaddress{\city{Sendai}, \state{Miyagi}, \country{Japan}}}

\affil[2]{\orgdiv{Neutron Science and Technology Center}, 
\orgname{Comprehensive Research Organization for Science and Society (CROSS)}, 
\orgaddress{\city{Tokai}, \state{Ibaraki}, \country{Japan}}}

\affil[3]{\orgdiv{Institute for Integrated Radiation and Nuclear Science}, 
\orgname{Kyoto University}, 
\orgaddress{\city{Kumatori}, \state{Osaka}, \country{Japan}}}

\affil[4]{\orgdiv{Department of Applied Physics and Physico-Informatics}, 
\orgname{Keio University}, 
\orgaddress{\city{Yokohama}, \state{Kanagawa}, \country{Japan}}}

\affil[5]{\orgdiv{Center for Science and Innovation in Spintronics (CSIS)}, 
\orgname{Tohoku University}, 
\orgaddress{\city{Sendai}, \state{Miyagi}, \country{Japan}}}

\abstract{
The spiral handedness of magnetic moments, referred to as chirality, gives rise to emergent electromagnetic phenomena in helimagnets.  
In insulating helimagnets, known as multiferroics, the cycloidal spin structure induces electric polarization by utilizing the inverse Dzyaloshinskii-Moriya mechanism. 
Spin-polarized neutron diffraction experiments, which directly probe circular spin arrangements, clearly demonstrated that an electric field controlled the chirality in multiferroic helimagnets. 
On the other hand, it was unclear until recently how the chirality could be controlled in metallic helimagnets where a large electric field cannot be applied, while the chirality control technique  in metallic helimagnets should enable the exploration of chirality-dependent spintronic functionalities. 
Recently, Jiang {\it et al.} succeeded in controlling the chirality of a spiral structure by the simultaneous application of a magnetic field and electric current in a metallic helimagnet, utilizing the nonreciprocal electronic transport as an indirect probe of chirality, highlighting the need for a neutron diffraction experiment that directly probes the chirality. 
Here, we directly demonstrate the chirality control in a metallic helimagnet YMn$_6$Sn$_6$ by means of spin-polarized neutron diffraction, which should give rise to a firm basis for the development of future helimagnetic spintronics.
}


\maketitle

\begin{figure}
    \centering
    \includegraphics[width=0.5\linewidth]{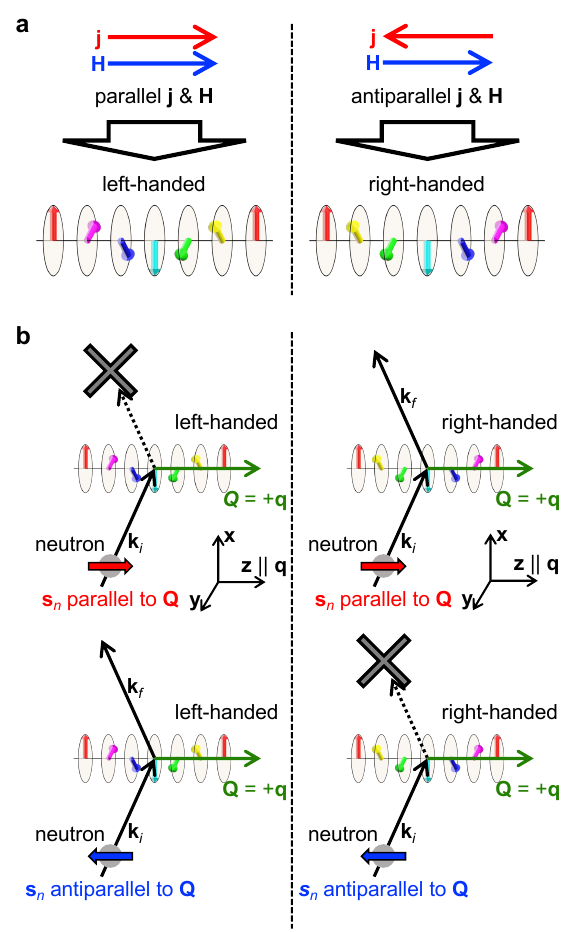}
    \caption{\label{fig:fig1}
    \textbf{Control and detection of helimagnetic chirality.}
    \textbf{a}, Schematic illustration of the chirality control in a metallic helimagnet by electric current $\mathbf{j}$ and magnetic field $\mathbf{H}$. 
    The left-handed chiral state is stabilized by applying $\mathbf{j}$ and $\mathbf{H}$ parallel to each other.
    The right-handed chiral state is, on the other hand, stabilized by applying antiparallel $\mathbf{j}$ and $\mathbf{H}$.
    \textbf{b}, Schematic illustration of the spin-polarized neutron diffraction from a helimagnetic structure.
    ${\mathbf q}$ is the magnetic propagation vector of the helimagnet,
    ${\mathbf s}_n$ is the incident neutron spin,
    ${\mathbf k}_i$ and ${\mathbf k}_f$ are the wave vectors of the incident and scattered neutrons, 
    and ${\mathbf Q} = {\mathbf k}_i - {\mathbf k}_f$ denotes the scattering vector.
    For the left-handed chiral helimagnetic structure,
    the magnetic Bragg reflection with ${\mathbf Q} = +{\mathbf q}$ vanishes when ${\mathbf s}_n$ is parallel to ${\mathbf Q}$ and is nonzero when ${\mathbf s}_n$ is antiparallel to ${\mathbf Q}$.
    For the right-handed chiral helimagnetic structure, on the other hand, 
    magnetic reflection is nonzero when ${\mathbf s}_n$ is parallel to ${\mathbf Q}$ and vanishes when ${\mathbf s}_n$ is antiparallel to ${\mathbf Q}$.
    }
\end{figure}

While magnetic memories currently utilize magnetization direction in ferromagnets, the stray field from the ferromagnet hinders further integration. 
To overcome this limitation, antiferromagnets with multiple distinguishable ground states, such as altermagnets, have attracted significant attention as promising candidates for next-generation magnetic memory\cite{MacDonald2011_AFMspintronics, Gomonay2014_AFMspintronics, Jungwirth2016_AFMspintronics, Baltz2018_AFMspintronics, Jungwirth2022_Altermag}.
Helimagnet, in which localized moments are spirally ordered, is also one such antiferromagnets\cite{Yoshimori1959}.
The right- and left-handed spiral magnetic states in helimagnets are degenerate when the crystal structure has inversion symmetry. 
Recent studies have demonstrated that the handedness degree of freedom, denoted as chirality, can be controlled by the application of a magnetic field and electric current (Fig. \ref{fig:fig1}\textbf{a})\cite{Jiang2020_MnP, Masuda2024_MnAu2, Yamaguchi2025_YMn6Sn6, Masuda2025_YMn6Sn6}.
Controlled chirality was detected by means of an indirect probing method of chirality, that is, nonreciprocal electronic transport. 
These studies motivated the exploration of helimagnetic functionalities, such as the sliding motion of magnetic structures \cite{Kimoto2025_sliding} and $p$-wave spin splitting\cite{Mayo2024_EuP3NET}.
Before proceeding with the further development of spintronic functionality in helimagnets, a direct and unambiguous demonstration of the chirality control should be performed.
In this study, by using spin-polarized neutron scattering technique, 
we have directly demonstrated the chirality control by a magnetic field and electric current in a metallic helimagnet YMn$_6$Sn$_6$.

Spin-polarized neutron diffraction is a powerful tool for probing the chirality in a helimagnet\cite{chatterji2005_neutron, Furrer2009_neutron}. 
When magnetic moments $\mathbf{S}\left(\mathbf{r}\right)$ form a helimagnetic structure as 
$\mathbf{S}\left(x,y,z\right) = 
S \cos \left(\chi qz\right) \hat{\mathbf{x}} +S \sin \left(\chi q z\right) \hat{\mathbf{y}}$,
the scattering cross sections of magnetic Bragg reflections with the scattering vector $\mathbf{Q}=\pm\mathbf{q}$ depend on the incident neutron spin $\mathbf{s}_n$ as\cite{Blume1963}
\begin{equation}
\label{eq1:CrossSection}
\left(\frac{d\sigma}{d\Omega}\right)^{\pm}=
A
\left\{
1+\left(\hat{\mathbf{Q}} \cdot \hat{\mathbf{q}}\right)^2 
\pm 2\chi\left(\hat{\mathbf{s}}_n \cdot \hat{\mathbf{Q}}\right)
\left(\hat{\mathbf{Q}} \cdot \hat{\mathbf{q}}\right)
\right\}.
\end{equation}
Here, $S$ is the magnitude of the magnetic moments,
$q$ is the magnitude of the magnetic propagation vector $\mathbf{q}=q\hat{\mathbf{z}}$ along the $z$ axis and 
$\hat{\mathbf{x}}, \hat{\mathbf{y}}, \hat{\mathbf{z}}$ are the unit vectors of the right-handed orthogonal coordinates.
$\chi=\pm1$ represents the chirality of the helimagnetic structure; 
$\chi=+1$ and $-1$ correspond to right-handed and left-handed helimagnets, respectively.
$A$ is a proportional constant determined by the magnetic structural factor and the hat symbols indicate the unit vectors.
The third term on the right hand side of Eq. \ref{eq1:CrossSection}, known as the chiral term, arises the scattering intensity depending on the chirality $\chi$, incident neutron spin $\mathbf{s}_n$ and scattering vector $\mathbf{Q}$.
This leads to extinction conditions of the scattering intensity for $\mathbf{Q}=\pm\mathbf{q}$ depending on the chirality and $\mathbf{s}_n$, as schematically depicted in Fig. \ref{fig:fig1}\textbf{b}.
For the left-handed chiral helimagnetic structure ($\chi=-1$), the scattering intensity vanishes when $\mathbf{s}_n$ is parallel to $\mathbf{Q}$ but it is nonzero when antiparallel. 
This relationship is reversed for the right-handed chiral helimagnetic structure  ($\chi=+1$).
By measuring the intensity of the magnetic reflections while flipping the incident neutron spin $\mathbf{s}_n$, the volume fraction of the two chiral states can be determined.
This method was utilized to demonstrate the chirality control in an insulating helimagnet by electric and magnetic fields in a pioneering work by Siratori et al.\cite{Siratori1980_ZnCr2Se4}
Subsequently, the correlation between chirality and electric polarization was demonstrated by this method in multiferroics\cite{Yamasaki2007_TbMnO3neutron, Seki2008_LiCu2O2, Sagayama2008_MnWO4, Nakajima2008_CuFeO2, Cabrera2009_Ni3V2O8, Fukunaga2009_TmMn2O5, Soda2009_CuCrO2, Finger2010_MnWO4, Tokura2010_Multiferroics, Wakimoto2013_YMn2O5, Ressouche2014_PolarizeNeutron}.
Here, we utilize this canonical method to demonstrate the chirality control in a metallic helimagnet YMn$_6$Sn$_6$.

\begin{figure}
    \centering
    \includegraphics[width=\linewidth]{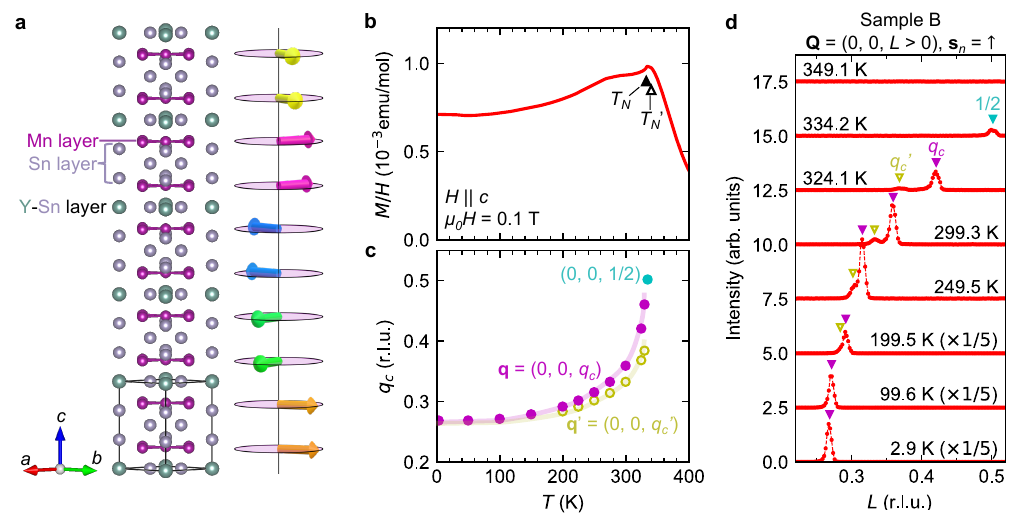}
    \caption{\label{fig:fig2}
    \textbf{Helimagnetic order in YMn$_6$Sn$_6$.}
    \textbf{a}, Crystal and  magnetic structures of YMn$_6$Sn$_6$ at 2 K\cite{Venturini1996_YMnSnneutron}.
    The arrows represent the directions of the Mn moments within the kagome plane.
    The magnetic structure corresponding to the predominant magnetic propagation vector $\mathbf{q}=(0,0,q_c)$ is shown.
    The crystal structure is drawn by VESTA\cite{VESTA}.
    \textbf{b}, Magnetic susceptibility $M/H$,
    which is obtained by the magnetization $M$ divided by magnetic field $H$,
    as a function of temperature $T$ at $\mu_0H=0.1$ T for $H\parallel c$. 
    Open and filled triangles denote the paramagnetic to antiferromagnetic transition temperature $T_N'$ and the antiferromagnetic to helimagnetic transition temperature $T_N$, respectively.
    This figure is reproduced from ref. \cite{Masuda2025_YMn6Sn6}.
    \textbf{c}, $c^*$ component of the magnetic propagation vector $\mathbf{q}=(0,0,q_c)$ and that of additional propagation vector $\mathbf{q}'=(0,0,q_c')$ as a function of temperature $T$.
    Solid curves are guide to the eye.
    \textbf{d}, $(0,0,L)$ profiles of the neutron scattering intensity at selected temperatures.
    The triangles denote the positions of the magnetic reflections.
    The dashed lines connect the marker points for clarity.
    The data for 199.5 K, 99.6 K and 2.9 K  are multiplied by 1/5.
    }
\end{figure}

YMn$_6$Sn$_6$ crystallizes in a layered centrosymmetric hexagonal crystal structure composed of Mn kagome layers, Sn layers, and Y-Sn composite layers, as shown in Fig. \ref{fig:fig2}\textbf{a}.
In magnetically ordered states, the Mn localized moments order ferromagnetically within a kagome layer parallel to the $ab$ plane\cite{Venturini1996_YMnSnneutron}. 
The layer magnetic moments exhibit antiferromagnetic ordering with a temperature-dependent propagation vector along the $c^\ast$ direction. 
Figure \ref{fig:fig2}\textbf{c} plots the $c^\ast$ component of the magnetic propagation vector $q_c$ obtained from the $(0,0,L)$ profiles of the neutron scattering intensities shown in Fig. \ref{fig:fig2}\textbf{d}.
Two successive antiferromagnetic transitions occur at $T_N' =$ 340 K and $T_N =$ 332 K\cite{Zhang2020_YMnSnneutron, Ghimire2020_YMnSnneutron, Dally2021_YMnSnneutron}.
The former shows a subtle shoulder-like feature and the latter a clear kink  in the temperature dependence of the magnetic susceptibility (Fig. \ref{fig:fig2}\textbf{b}).
Between $T_N$ and $T_N'$, the Mn moments exhibit a collinear antiferromagnetic structure with the propagation vector $(0,0,1/2)$\cite{Zhang2020_YMnSnneutron}.
Below $T_N$, an incommensurate helimagnetic structure appears, where the Mn moments in two Mn kagome layers sandwiching the Sn layer are nearly ferromagnetically aligned and are rotated along the $c$ axis across the Y-Sn layer, as depicted in Fig. \ref{fig:fig2}\textbf{a}\cite{Venturini1996_YMnSnneutron}.
It should be noted that in addition to the predominant magnetic reflection at $\mathbf{q}=(0,0,q_c)$, there is an additional magnetic modulation with $\mathbf{q}'=(0,0,q_c')$.
$q_{c}$ and $q_{c}'$ vary smoothly with their difference decreasing on cooling.
The origin of the additional peak is discussed later.

\begin{figure}
    \centering
    \includegraphics[width=.9\linewidth]{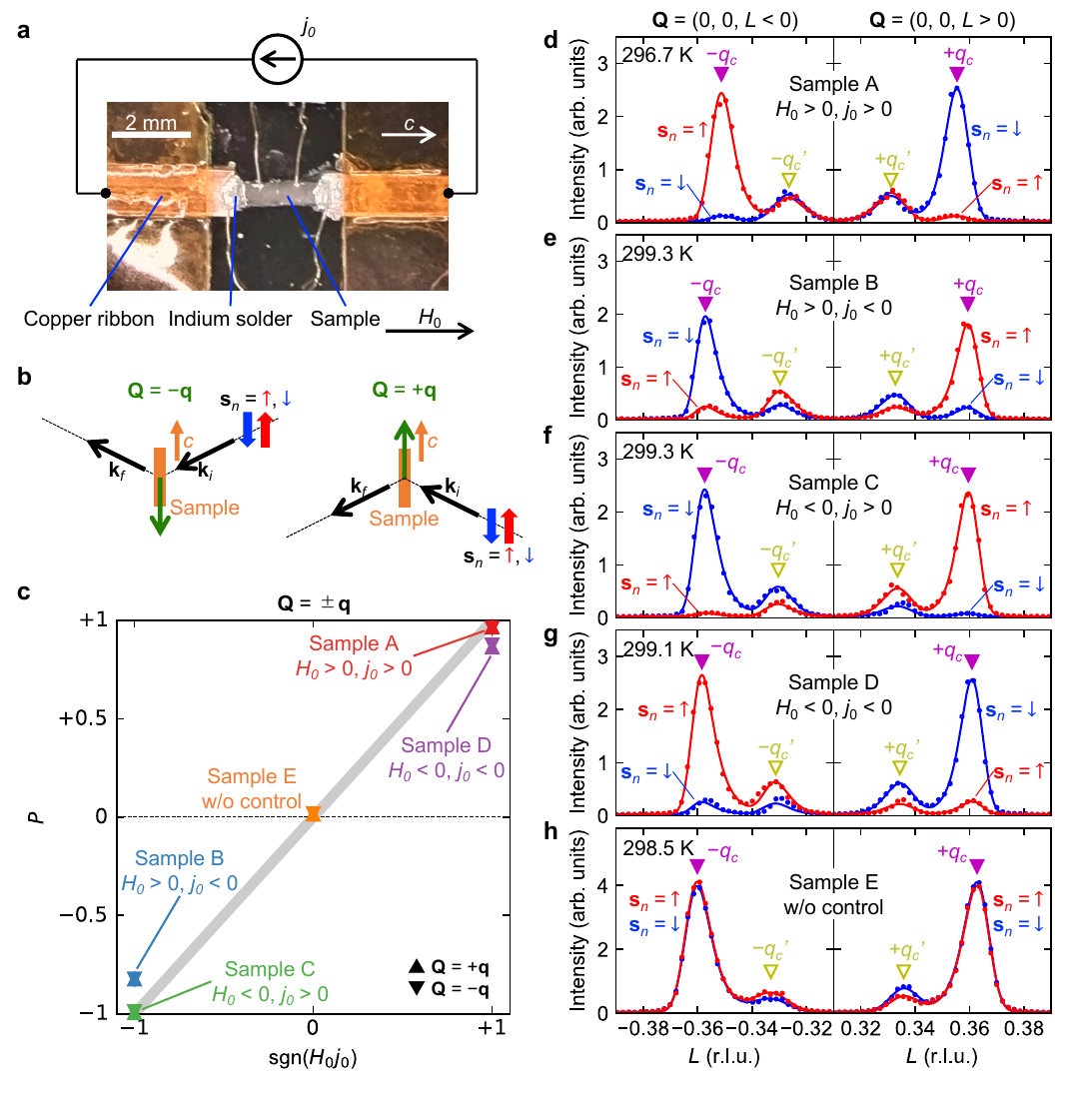}
    \caption{\label{fig:fig3}
    \textbf{Demonstration of helimagnetic chirality control.}
    \textbf{a}, Image of the setup for chirality control by electric current and magnetic field.
    A YMn$_6$Sn$_6$ single-crystal sample was fixed on a substrate and connected to copper ribbons using indium solder.
    Positive and negative electric currents $j_0$ and magnetic fields $H_0$ were applied along the $c$ axis.
    \textbf{b}, Top-view schematic illustration of the setup for spin-polarized neutron diffraction measurements.
    The scattering vector ${\mathbf Q}=\pm{\mathbf q}$ and incident neutron spin ${\mathbf s}_n$ were parallel or antiparallel to the $c$ axis.
    See Extended Data Fig. \ref{fig:figS_setup} for more details.
    \textbf{c}, Excess factor
    $P=
    \mp\frac{I_{\pm\mathbf{q},\uparrow}-I_{\pm\mathbf{q},\downarrow}}{I_{\pm\mathbf{q},\uparrow}+I_{\pm\mathbf{q},\downarrow}} \frac{1}{p_n}$
    as a function of ${\rm sgn}(H_0j_0)$.
    $I_{\pm\mathbf{q},\uparrow}$ and $I_{\pm\mathbf{q},\downarrow}$ are the integrated intensities of $\mathbf{Q}=\pm\mathbf{q}$ magnetic reflections for ${\mathbf s}_n=\uparrow$ and ${\mathbf s}_n=\downarrow$, respectively,
    and $p_n=0.955(2)$ is the spin polarization of the incident neutron beam\cite{Morikawa2024_TAIKAN}.
    \textbf{d-h}, $L$-scan profiles of the neutron scattering intensities around the 
    $\mathbf{Q}=\pm\mathbf{q}$
    magnetic reflections for Samples A-E at room temperature.
    The triangles denote the positions of the magnetic reflections.
    Solid curves are fit to a phenomenological function (see Methods).
    }
\end{figure}

Then, let us demonstrate the chirality control using spin-polarized neutron diffraction. 
For neutron diffraction experiments, we need a single crystal larger than a millimeter, but for such a large crystal it is difficult to apply a large electric current density required for the chirality control. 
To overcome this difficulty, we developed a setup for applying large electric currents in a superconducting magnet (Fig. \ref{fig:fig3}\textbf{a}). 
We reduced the resistance of the wires and contacts by using thick copper ribbons and indium solders.
Since $T_N$ is above room temperature, the controlled chirality should be preserved even if the sample is removed from the cryostat. 
We controlled the chirality of several YMn$_6$Sn$_6$ crystals at the Institute for  Materials Research, Tohoku University, and brought them to MLF, J-PARC for neutron diffraction experiments. 
For the chirality control, we applied a magnetic field $\mu_0H_0=\pm 7$ T and a large dc electric current of $I_0 \approx \pm5$ A along the $c$-axis in parallel or antiparallel to each other, and then swept the magnetic field from $H_0$ to 0 T.
The electric current of $I_0 \approx 5$ A corresponds to a current density of $j_0 \approx \pm 5\times 10^7$ A/m$^2$, which is above the threshold current for chirality control\cite{Masuda2025_YMn6Sn6}.
The details of the sample preparation and chirality control procedure are provided in the Methods section.
We prepared five samples for the spin-polarized neutron diffraction measurements as summarized in Extended Data Tab. \ref{tab:SampleList}. 
The chiralities of four samples were controlled using positive and negative magnetic fields and electric currents: $j_0>0$ and $H_0>0$ for Sample A, $j_0<0$ and $H_0>0$ for Sample B, $j_0>0$ and $H_0<0$ for Sample C, and $j_0<0$ and $H_0<0$ for Sample D. 
The chirality of Sample E was left uncontrolled. 
According to the measurement of nonreciprocal electronic transport, the chiralities of Samples A and D are controlled to be the same one and those for Samples B and C are opposite\cite{Jiang2020_MnP,Masuda2024_MnAu2,Masuda2025_YMn6Sn6}. 
In Sample E, domains with two chiral states should be equally distributed. 
To directly probe these chiral states, we performed spin-polarized neutron scattering measurements on Samples A-E.
As described above, the neutron spin $\mathbf{s}_n$ dependence of the scattering intensity probes the chirality. 
In our experimental setup (Fig. \ref{fig:fig3}\textbf{b}), the samples were mounted so that the $c$-axis was within the horizontal scattering plane, in order to observe the $\mathbf{Q}=\pm\mathbf{q}=(0,0,\pm q_c)$ magnetic reflections, and the incident neutron spin $\mathbf{s}_n$ was aligned parallel ($\mathbf{s}_n=\uparrow$) or antiparallel ($\mathbf{s}_n=\downarrow$) to the $c$ axis.
Details of the measurement configuration are given in Methods section and Extended Data Fig. \ref{fig:figS_setup}.
As mentioned above, for the left-handed state, the $\mathbf{Q}=+\mathbf{q}$ scattering intensity should vanish when $\mathbf{s}_n=\uparrow$ and for the right-handed state when $\mathbf{s}_n=\downarrow$. 
These relations are reversed in the case of $\mathbf{Q}=-\mathbf{q}$ scattering. 
Let us discuss the experimental results keeping these relations in mind.
Figure \ref{fig:fig3}\textbf{d} shows the $L$-scan profiles of spin-polarized neutron scattering intensity around the $\mathbf{Q}=\pm\mathbf{q}$ magnetic reflections for Sample A measured at 296.7 K.
The intensity of the $\mathbf{Q}=+\mathbf{q}$ reflection was approximately 25 times larger for $\mathbf{s}_n=\downarrow$ than that for $\mathbf{s}_n=\uparrow$.
As for the  $\mathbf{Q}=-\mathbf{q}$ reflection, conversely,
intensity for $\mathbf{s}_n=\uparrow$ was significantly larger than that for $\mathbf{s}_n=\downarrow$.
These neutron spin dependences suggest that the majority of the sample volume is controlled to the left-handed chiral state. 
As shown in Figs. \ref{fig:fig3}\textbf{e, f, g}, the neutron spin dependences for Samples B and C were opposite from that for Sample A, while it was similar for Sample D.
In the case of Sample E, the $\mathbf{s}_n$ dependence was negligible. 
These results are quite consistent with the expected chiral states controlled by the magnetic field and electric current. 
While only the relative sign of chirality was identified in the previous reports, the spin-polarized neutron diffraction detected the absolute handedness; 
the left-handed chirality is stabilized by the parallel electric current and magnetic field, and the right-handed chirality is under antiparallel field and current conditions.

The volume fraction of controlled chirality can be evaluated based on the neutron spin dependence of the  $\mathbf{Q}=\pm\mathbf{q}$ reflection by using the excess factor 
\begin{equation}
\label{eq:ExcessFactor}
P = 
\mp \frac{I_{\pm\mathbf{q},\uparrow}-I_{\pm\mathbf{q},\downarrow}}{I_{\pm\mathbf{q},\uparrow}+I_{\pm\mathbf{q},\downarrow}} \frac{1}{p_n},
\end{equation}
where 
$I_{\pm\mathbf{q},\uparrow}$ ($I_{\pm\mathbf{q},\downarrow}$) is the integrated reflection intensities for $\mathbf{Q}=\pm\mathbf{q}$ and incident neutron spin $\mathbf{s}_n=\uparrow$ ($\mathbf{s}_n=\downarrow$) calculated by fitting to a phenomenological function, and $p_n=0.955(2)$\cite{Morikawa2024_TAIKAN} is the polarization of the incident neutron spin (see Methods).
$P=+1$ corresponds to the fully polarized left-handed chiral state,
$P=-1$ to the fully polarized right-handed chiral state, 
and $P=0$ to a multidomain state where left- and right-handed domains are equally distributed.
In Fig. \ref{fig:fig3}\textbf{c}, $P$ for Samples A-E is plotted against ${\rm sgn}(H_0j_0)$, where $H_0$, $j_0$, and $\rm{sgn}(x)$ are magnetic field, electric current used for the chirality control and the sign function, respectively. 
This figure clearly shows that the chirality domains were effectively controlled depending on ${\rm sgn}(H_0j_0)$.
Magnitude of $P$ for Samples A-D ranges from 0.823(5) to 0.992(5) depending on the sample.
Imperfections in chiral domain control and its sample dependence may arise from the nonuniform electric current distribution near the electrical contacts.

For the additional magnetic peak at $\mathbf{Q}=\pm \mathbf{q}'$, the neutron spin dependence was weak and not systematic. 
The excess factor $P$ for $\mathbf{Q}=\pm \mathbf{q}'$ does not depend on ${\rm sgn}(H_0j_0)$ as shown in Extended Data Fig. \ref{fig:figS_P_q1}. 
While a previous paper suggested that the origin of the coexisting two propagation vectors seemed to be a long-range modulation of the helimagnetic structure\cite{Venturini1996_YMnSnneutron}, 
the present observation instead indicates the possibility of an origin totally independent of the main helimagnetic structure such as phase separation.

\begin{figure}
    \centering
    \includegraphics[width=.5\linewidth]{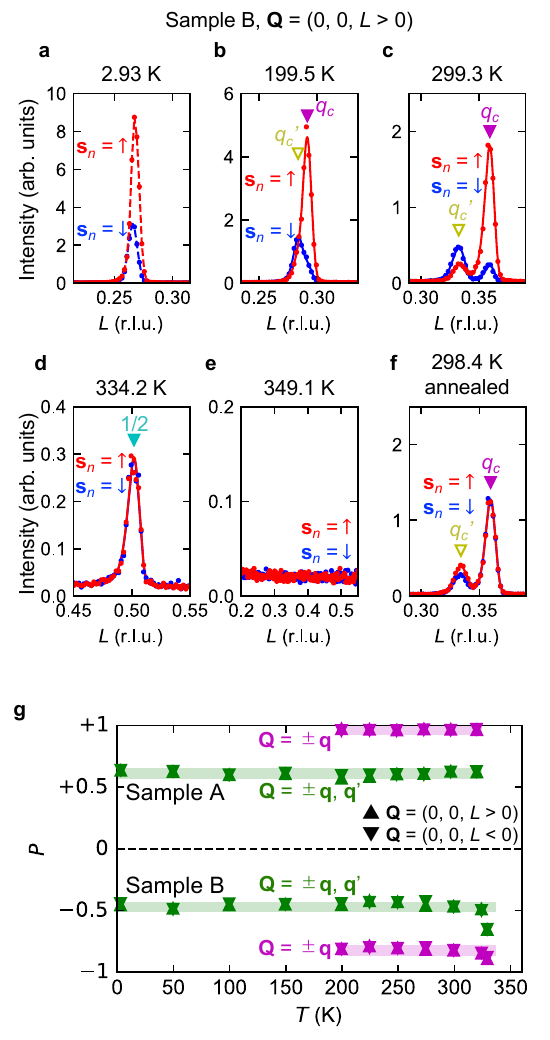}
    \caption{\label{fig:fig4}
    \textbf{Neutron spin dependence of scattering intensity at various temperatures.}
    \textbf{a-e}, $L$-scan profiles of neutron scattering intensities for sample B at selected temperatures.
    Data in \textbf{f} were measured after the annealing procedure, \textit{that is}, heating above 400 K and cooling down to room temperature.
    The solid curves in \textbf{b-f} are fit to a phenomenological function similar to Figs. \ref{fig:fig3}\textbf{d-h}, while the dashed lines in \textbf{a} merely connect the data points.
    \textbf{g}, Excess factor $P$
    for the $\mathbf{Q}=\pm\mathbf{q}$ reflections (purple) and the total excess factor for both $\mathbf{Q}=\pm\mathbf{q}$ and $\mathbf{Q}=\pm\mathbf{q}'$ reflections (green) for Samples A and B.
    Solid lines are guide to the eye.
    }
\end{figure}

The $L$-scan profiles of the spin-polarized neutron scattering intensity for Sample B at the selected temperatures are shown in Figs. \ref{fig:fig4}\textbf{a-e}.
For this sample, the neutron spin dependence of the additional peak ($\mathbf{Q}=\pm\mathbf{q}'$) is weak and opposite to that of the main peak ($\mathbf{Q}=\pm\mathbf{q}$). 
As the temperature is lowered from the room temperature,
the peaks at $\mathbf{Q}=\pm\mathbf{q}$ and at $\mathbf{Q}=\pm\mathbf{q}'$ are gradually merged, and the neutron spin polarization seems averaged for the merged peak. 
Figure \ref{fig:fig4}\textbf{g} shows the temperature dependence of the excess factor $P$ for the $\mathbf{Q}=\pm\mathbf{q}$ main reflection, as well as the total excess factor for both the $\mathbf{Q}=\pm\mathbf{q}$ and $\mathbf{Q}=\pm\mathbf{q}'$ reflections observed in Samples A and B. 
The total excess factor was estimated by numerical integration through two peaks. 
These quantities are almost independent of the temperature, and the magnitudes of the total excess factor is smaller because of the weak spin dependence of additional $\mathbf{Q}=\pm\mathbf{q}'$ scattering peak.

At 334.2 K in the colinear AFM phase with the commensurate propagation vector $(0,0,1/2)$,
the reflection intensity becomes independent of $\mathbf{s}_n$ as shown in Fig. \ref{fig:fig4}\textbf{d}, which is consistent with the absence of chirality in the colinear antiferromagnetic state.
In the paramagnetic state above $T_N'$, magnetic reflection was no longer observed, as shown in Fig. \ref{fig:fig4}\textbf{e}.
Importantly, we observed the disappearance of chirality polarization after annealing Sample B at a high temperature.
After heating the Sample B above 400 K and cooling it down to the room temperature,
the reflection intensities are almost independent of $\mathbf{s}_n$ as shown in Fig. \ref{fig:fig4}\textbf{f}, indicating that the two chiral domains becomes equally distributed after the annealing process. 
The disappearance of the neutron spin dependence further ensures that the neutron spin dependence shown in Figs. \ref{fig:fig3}\textbf{d-g} probes the chirality controlled by the magnetic field and electric current.

In summary, we directly demonstrated chirality control by a magnetic field and an electric current using spin-polarized neutron diffraction. 
While it was previously demonstrated by means of the nonreciprocal electronic transport, the spin-polarized neutron diffraction can directly probe the magnetic structure free from any experimental artifact, such as Joule heating for the nonreciprocal electronic transport, and therefore provides definite evidence of chirality control. 
In addition, while the nonreciprocal transport identifies only the relative sign of chirality, spin-polarized neutron diffraction provides the absolute sign of chirality; we have found that a parallel magnetic field and electric current stabilize the left-handed chirality, whereas an antiparallel magnetic field and electric current stabilize the right-handed chirality. 
This information should be useful for the examination of the theoretical mechanisms of chirality-related spintronic phenomena. 
Thus, this observation may serve as a cornerstone for the future development of helimagnet-based spintronics.

\section*{Methods}\label{sec:methods}
\subsection*{Crystal growth}
Single crystals of YMn$_6$Sn$_6$ were grown using the Sn self-flux method. 
High-purity chunks of Y (99.9 \%), Mn ($>$99.9 \%), and Sn (99.99 \%) were mixed with the ratio Y : Mn : Sn = 1 : 6 : 30 
and placed in an alumina crucible. 
The crucible was then sealed in an evacuated quartz tube and heated at 1000 ${\rm ^\circ C}$ for 20 hours, 
followed by slow cooling to 600 ${\rm ^\circ C}$ at the rate of 1 $^\circ$C/hour 
where the excess Sn flux was decanted using a centrifuge.
Hexagonal plate-like YMn$_6$Sn$_6$ single crystals with maximum sizes of $5\times5\times2$ mm$^3$ were obtained. 

\subsection*{Chirality control}
The YMn$_6$Sn$_6$ crystals were cut into bar-shaped pieces with a typical size of $0.2\times0.6\times2.2$ mm$^3$ and a typical mass of 2 mg.
The sample was then fixed onto a surface-oxidized Si substrate and connected to copper ribbons with a thickness of 0.1 mm using indium solder, as shown in Fig. \ref{fig:fig3}\textbf{a}.
A good electrical contact with a typical contact resistance of 10 $\rm{m\Omega}$ was achieved.
Voltage contacts for the four-probe resistivity measurements were made using silver paste.
The application of a magnetic field and electric current for the chirality control was conducted in a superconducting magnet installed at the Institute for Materials Research, Tohoku University.
While sweeping a magnetic field from $\mu_0H_0=\pm7$ T to 0 T, 
a large dc electric current $I_0$ with a typical magnitude of 5 A was applied through the copper ribbons.
The magnitude of the corresponding current density $j_0$ is typically $5\times10^7$ A/m$^2$.
While the temperature of the sample stage was maintained at 150 K, 
Joule heating due to the large dc electric current $I_0$ appears to increase the sample temperature estimated by the sample resistance up to approximately 300 K, as shown in Extended Data Fig. \ref{fig:figS1_Idc2R}.
Then, we turned off $I_0$, slowly warmed the sample to 300 K while carefully avoiding overshooting, and removed it from the superconducting magnet.
Sample size, mass, $H_0$, $I_0$ and $j_0$ for Samples A-E are summarized in Extended Data Tab. \ref{tab:SampleList}.

\subsection*{Spin-polarized neutron scattering}
After the chirality control, samples were transfered from the Si substrates to Al sample holders for the neutron scattering measurements.
Spin-polarized neutron scattering measurements were performed by using a time-of-flight (TOF) small- and wide-angle neutron scattering instrument TAIKAN constructed at the BL15 of the Materials and Life Science Experimental Facility (MLF), Japan Proton Accelerator Research Complex (J-PARC), Japan\cite{Takata2015_TAIKAN}.
An incident neutron beam with a wavelength ranging from 0.7 to 7.7 \AA\ was exposed on the sample. 
Sketch of the measurement configuration is shown in Extended Data Fig. \ref{fig:figS_setup}.
Sample was attached to the goniometer so that the $c$ direction point rightward from the viewpoint of the incident beam, where the positive $c$ direction was defined parallel to the positive magnetic field direction of the superconducting magnet used for the chirality control (Fig. \ref{fig:fig3}\textbf{a}).
The polarized neutron beam was provided by a multichannel supermirror installed in the upstream optics section of the beamline\cite{Shinohara2009_polarizer}.
The incident neutron polarization $p_n=0.955(2)$ for the typical neutron wavelength of 4.4 \AA\ \cite{Morikawa2024_TAIKAN} was used for the evaluation of excess factor in Eq. \ref{eq:ExcessFactor}, where $p_n$ is defined so that $p_n=$1 and 0 corresponds to the perfectly polarized and unpolarized beam, respectively.
The neutron spin polarity was reversed with a gradient radio frequency neutron spin flipper\cite{Weifurter1989_SpinFlipper} placed after the polarizer in the upstream optics section, where the flipping efficiency of the flipper was extremely close to 1 for incident neutrons with $\lambda \ge$2.0 \AA.
The direction of the spin polarization axis was aligned parallel to the $c$ axis using guide fields of $\sim1\times10^{-2}$ T, as depicted in Extended Data Fig. \ref{fig:figS_setup}.
Spin flipper ON and OFF correspond to the incident neutron spin parallel ($\mathbf{s}_n=\uparrow$) and antiparallel ($\mathbf{s}_n=\downarrow$) to the $c$ axis, respectively.

\subsection*{Analysis of the neutron diffraction peak profiles}
The observed neutron intensities were processed using Utsusemi software\cite{Utsusemi}.
The $L$-scan profiles of the observed magnetic diffraction were analyzed using original python codes, as shown in Figs. \ref{fig:fig3}\textbf{d-h}, \ref{fig:fig4}\textbf{a-f}.
The peak profiles exhibited an asymmetric shape with a long exponential tail extending toward the low-$Q$, long-wavelength side.
This feature is known to arises from TOF instrumentation using pulsed neutron spallation sources\cite{VonDreele1982_TOFprofile, GSAS}.
To fit the peak profiles, we utilized a phenomenological function
\[ H\left(Q\right) = \frac{\alpha}{2} e^{u-y^2} {\rm erfcx} (y), \]
where 
$u = \frac{1}{2}\alpha^2\sigma^2 + \alpha \Delta Q$,
$y = \frac{\alpha\sigma^2 - \Delta Q }{\sqrt{2\sigma^2}}$,
$\Delta Q = Q - Q_0 $,
and erfcx is the scaled complementary error function.
$Q_0$, $\alpha$, and $\sigma$ were used as the fitting parameters describing the peak position and shape.
This function is obtained by convolving a one-sided exponential function
\begin{align*}
E(Q') 
&= \alpha e^{+\alpha Q'} &\left( Q'<0 \right) \\
&= 0 &\left( Q'>0 \right) 
\end{align*}
and a Gaussian function
\[
G(\Delta Q -Q') = 
\frac{1}{\sqrt{2 \pi \sigma^2}} 
\exp \left[-\frac{(\Delta Q -Q')^2} { 2 \sigma^2} \right]
\]
as
\[
H\left( Q\right) = \int G(\Delta Q - Q') E(Q') dQ'.
\]
This function exhibits slow exponential decay at the low-$Q$ ($Q - Q_0 <0$) side and fast Gaussian-like decay at the high-$Q$ ($Q - Q_0 <0$) side, and appears to reproduce the observed peak profiles phenomenologically well.
The peak positions (triangles in Figs. \ref{fig:fig2}\textbf{d}, \ref{fig:fig3}\textbf{d-h} and \ref{fig:fig4}\textbf{a-f}) are defined by the maximum of the peak profile.

\bibliography{YMn6Sn6TAIKAN}

\section*{Acknowledgments}
The authors acknowledge N. Jiang for his efforts on similar experiments using another metallic helimagnet MnP, which helped us in developing the setup for the chirality control using a large electric current.
The authors also thank Y. Kawamura and T. Morikawa for their help during neutron scattering experiments.
This work was partially supported by JSPS (KAKENHI Nos. 
JP22H05145, 
JP23K13654, 
JP24H01638, 
JP24H00189,  
JP24K00572, 
JP25K01489) 
 and JST (PRESTO No. JPMJPR245A and 
FOREST No. JPMJFR202V). 
Neutron scattering experiments at the Materials and Life Science Experimental Facility of J-PARC were performed under user program proposal No. 2024B0119.

\clearpage

\clearpage
\section*{Extended Data}\label{sec:ext}
\renewcommand{\figurename}{Extended Data Fig.}
\renewcommand{\tablename}{Extended Data Tab.}
\setcounter{figure}{0} 

\begin{table}[h]
    \centering
    \begin{tabular}{ccccccc}
        \hline \hline
        Sample No. & Size (mm$^3$) & Mass (mg) 
        & $\mu_0H_0$ (T) & $I_0$ (A) & $j_0$ ($10^7$ A/m$^2$) & ${\rm sgn}(H_0j_0)$ \\
        \hline 
        A & $0.19\times0.60\times2.2$ & 2.29 
            & $+7.0$ & $+5.65$ & $+4.9$ & $+1$ \\
        B & $0.16\times0.57\times2.2$ & 1.77 
            & $+7.0$ & $-4.60$ & $-5.1$ & $-1$ \\
        C & $0.18\times0.58\times2.2$ & 1.99 
            & $-7.0$ & $+4.90$ & $+4.8$ & $-1$ \\
        D & $0.17\times0.62\times2.2$ & 2.01 
            & $-7.0$ & $-5.25$ & $-4.9$ & $+1$ \\
        E & $1.6 \times0.67\times2.2$ & 17.62 
            & \multicolumn{3}{c}{(uncontrolled)} & 0 \\
        \hline \hline
    \end{tabular}
    \caption{
    \textbf{List of the samples used in the present experiments.}
    The sample size, sample mass, magnetic field $H_0$, electric current $I_0$ and current density $j_0$ used for the field-sweep chirality control, and ${\rm sgn}(H_0j_0)$ for Samples A-E.
    }
    \label{tab:SampleList}
\end{table}

\begin{figure}[h]
    \centering
    \includegraphics[width=0.5\linewidth]{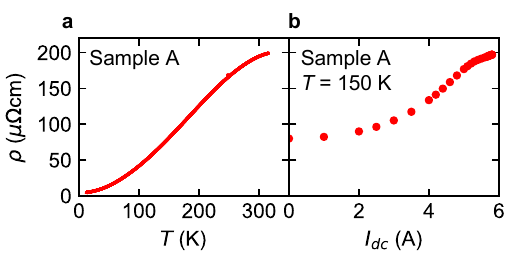}
    \caption{
    \textbf{Estimation of Joule heating.}
    \textbf{a}, Resistivity of Sample A as a function of temperature.
    \textbf{b}, Resistivity as a function of applied dc electric current $I_{dc}$.
    The resistivity was measured using a lock-in technique with a superimposed ac current of 10 mA.
    The temperature of the sample stage was maintained at 150 K.
    A dc current of $I_0=+5.65$ A appears to heat the sample up to approximately 300 K.
    }
    \label{fig:figS1_Idc2R}
\end{figure}

\begin{figure}
    \centering
    \includegraphics[width=1\linewidth]{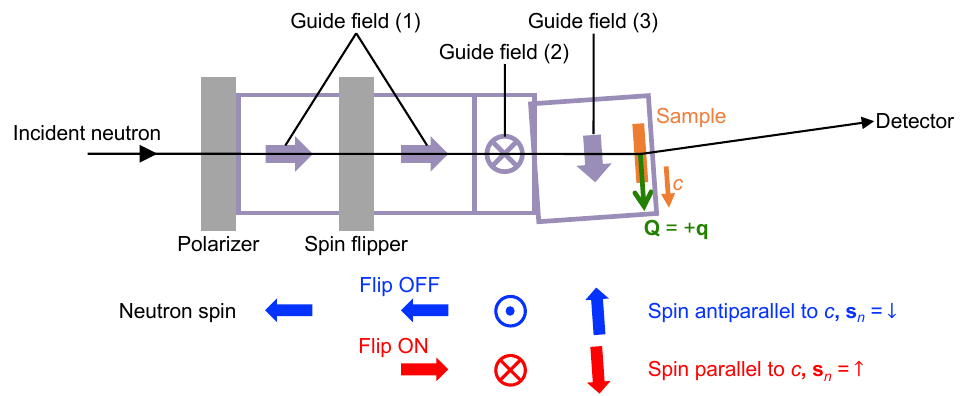}
    \caption{
    \textbf{Sketch of the spin-polarized neutron scattering measurement configuration.}
    Top-view sketch of the measurement configuration for the spin-polarized neutron scattering experiments in TAIKAN.
    Sample was attached to the goniometer so that the $c$ axis point rightward from the viewpoint of the incident beam, where the $c$ direction was defined by the positive magnetic field direction of the superconducting magnet used for the chirality control (Fig. \ref{fig:fig3}\textbf{a}).
    Spin-polarized incident neutron beam was provided by using spin polarizer and spin flipper installed in the upstream optics section of the beam line.
    The direction of the spin polarization axis was aligned along the $c$ axis using guide fields of $\sim1\times10^{-2}$ T.
    Guide field (3) was produced by permanent magnets fixed on the goniometer so that it is always parallel to the $c$ axis.
    Bottom panel shows the neutron spin directions through the beam line.
    Incident neutron spin was parallel ($\mathbf{s}_n=\uparrow$) and antiparallel ($\mathbf{s}_n=\downarrow$) to the $c$ axis when the spin flipper is ON and OFF, respectively.
    }
    \label{fig:figS_setup}
\end{figure}

\begin{figure}
    \centering
    \includegraphics[width=0.5\linewidth]{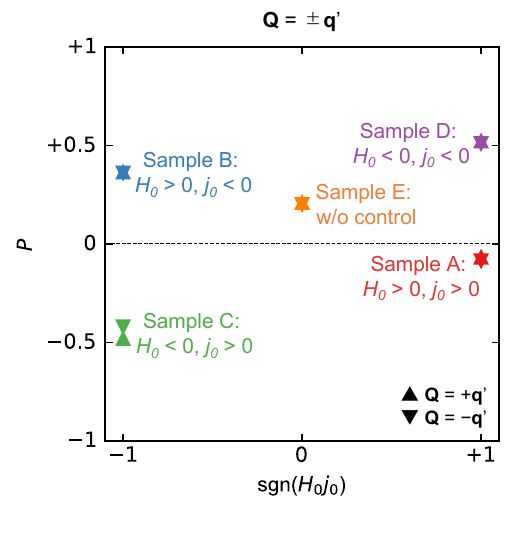}
    \caption{
    \textbf{Excess factor $P$ for $\pm\mathbf{q}'$ reflections.}
    Excess factor for the $\mathbf{Q}=\pm\mathbf{q}'$ magnetic reflections 
    $P = \mp \frac{I_{\pm\mathbf{q}',\uparrow}-I_{\pm\mathbf{q}',\downarrow}}{I_{\pm\mathbf{q}',\uparrow}+I_{\pm\mathbf{q}',\downarrow}} \frac{1}{p_n}$
    for Samples A-E, plotted against ${\rm sgn}(H_0j_0)$.
    $P$ does not depend on ${\rm sgn}(H_0j_0)$.
    }
    \label{fig:figS_P_q1}
\end{figure}

\begin{figure}
    \centering
    \includegraphics[width=.5\linewidth]{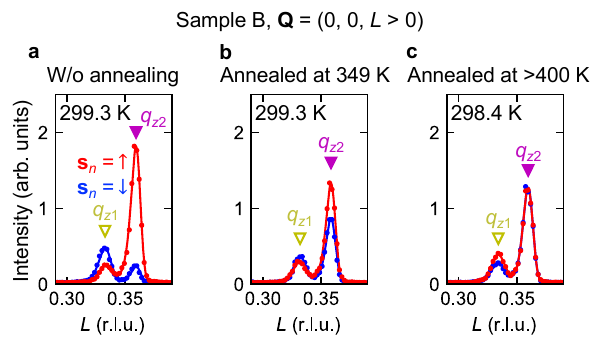}
    \caption{
    \textbf{Annealing temperature dependence of chirality.}
    $L$-scan profiles of the neutron scattering intensities around the $\mathbf{Q}=+\mathbf{q}$ magnetic reflection for Sample B at approximately 300 K 
    measured just after the chirality control (\textbf{a}), 
    measured after subsequent annealing at 349 K (\textbf{b}), 
    and measured after subsequent annealing at above 400 K (\textbf{c}).
    \textbf{a} and \textbf{c} are reproduced from Figs. \ref{fig:fig4}\textbf{a} and \ref{fig:fig4}\textbf{f}, respectively.
    Annealing at 349 K weakened the chiral domain polarization, but was not sufficient to fully annihilate it.
    }
    \label{fig:figS_anneal350K}
\end{figure}

\begin{figure}
    \centering
    \includegraphics[width=0.5\linewidth]{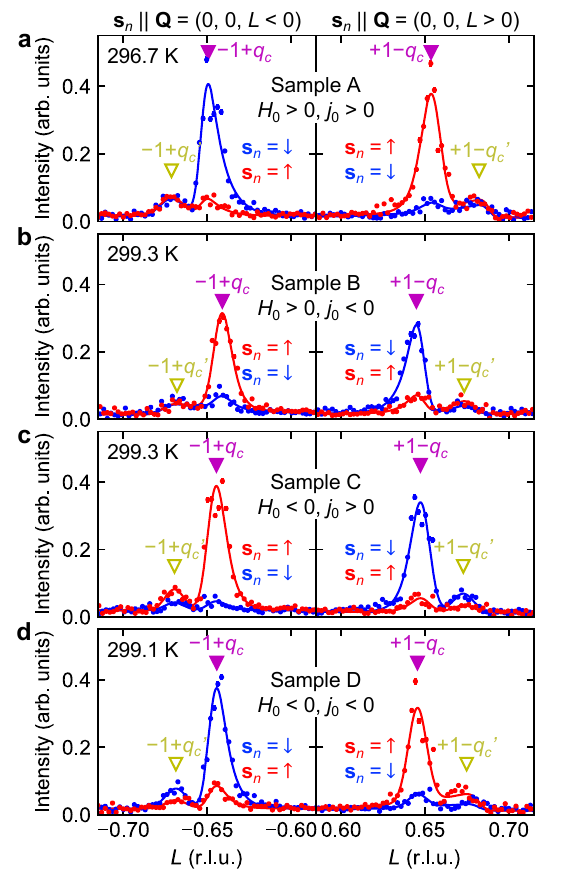}
    \caption{
    \textbf{Neutron spin dependence for $\mathbf{Q}=\left(0,0,\pm\left(1-q_c\right)\right)$ magnetic reflections.}
    $L$-scan profiles of the neutron scattering intensities around the $\mathbf{Q}=\left(0,0,\pm\left(1-q_c\right)\right)$ magnetic reflections for Samples A-D at approximately 300 K.
    Solid curves are fit to a phenomenological function similarly to Figs. \ref{fig:fig3}\textbf{d-h}.
    Incident neutron spin dependence of the $\mathbf{Q}=\left(0,0,+1-q_c\right)$ ($\mathbf{Q}=\left(0,0,-1+q_c\right)$) magnetic reflection is the same as that for the $\mathbf{Q}=\left(0,0,-q_c\right)$ ($\mathbf{Q}=\left(0,0,+q_c\right)$) reflection shown in Figs. \ref{fig:fig3}\textbf{d-g},
    which is consistent with the theoretically expected spin dependence\cite{Blume1963}.
    }
    \label{fig:figS_Q1mq_line}
\end{figure}

\end{document}